\documentclass[twocolumn,showpacs,preprintnumbers,amsmath,amssymb]{revtex4}

\begin{document}

\preprint{}

\title{Reply to "Comment on
    'Scalar-tensor gravity coupled to\\ a global monopole
    and flat rotation curves' "}

\author{Tae Hoon Lee
and Byung Joo Lee}
\affiliation{Department of Physics, Soongsil
University, Seoul 156-743, Korea}

\date{\today}% It is always \today, today,

\begin{abstract}
In Brans-Dicke theory of gravity we explain how the extra constant
value in the formula for rotation velocities of stars in a
galactic halo can be obtained due to the global monopole field. We
argue on a few points of the preceding Comment and discuss
improvement of our model.
\end{abstract}

\maketitle

%%%%%%%%%%%%%%%%%%%%%%%%%%%%%%%%%%%%%%%%%%%%%%%%%%%%%%%%
\section{Introduction}

In Ref. \cite{TB} we considered a global monopole(GM) \cite{vi} as
a candidate for galactic dark matter in Brans-Dicke(BD) theory of
gravity \cite{BD}. Within the weak gravity approximation we solved
the equations of metric tensor fields and BD scalar field coupled
to the GM, and determined the asymptotic structure of a galactic
spacetime at a distance $r$ from the monopole center. A metric
component in the so-called physical frame \cite{Bezerra} was given
by
\begin{equation}
\tilde{g}_{tt}(r) \simeq  -1- 2 (v_{rot}^{(0)})^2{\rm
ln}({r}/{r_i})+ 2G\int d^3 r' {\rho(r')}/\Delta r,
\end{equation}
with $\Delta r={\mid\vec{r}-\vec{r'}\mid}$, $G=G_*
(1+\alpha_0^2)$, and a ordinary matter density $\rho(r)$. From the
geodesic equation in the spacetime at the center of which there is
the monopole, we obtained a formula for the rotation velocity of
stars in the galactic halo, which contains an extra value
$v_{rot}^{(0)}$ in addition to the other known term
\begin{equation}
v^2_{rot} (r)= (v_{rot}^{(0)})^2+ {G M_{in} (r)}/{ r},
\end{equation}
with a mass $M_{in}(r)=4\pi\int_0^r dr' {r'}^2
\rho(r') $ of a sphere of radius $r$, and we discussed its
possible relation to flat rotation curve(RC)s in spiral galaxies.

In the preceding Comment \cite{Salucci}, Salucci and Gentile
suggested that the paradigm of flat RCs of spiral galaxies
\cite{FRC}, which was constructed about twenty years ago, has no
observational support now and that the notion of flat RCs might be
superseded by universal RCs \cite{URC}. The universal RC that they
advocated seems to be a great improvement on the flat RC in spiral
galaxies. However, the notion of flat RCs is supposed to remain
valid as a first approximation for large $r$ on RCs of some spiral
galaxies, and therefore we can not agree on their argument that
the paradigm of the flat RCs was dismissed.

In Sec. II, we explain how the extra constant value
$v_{rot}^{(0)}$ of the rotation velocity in the galactic halo
region can be generated by GM in BD theory of gravity, and we
argue upon the preceding Comment \cite{Salucci}. Section III
includes a summary and discussion.

%%%%%%%%%%%%%%%%%%%%%%%%%%%%%%%%%%%%%%%%%%%%%%%%%%%%%%%%%%%%%%%%%%%%%%%%
\section{BD gravity coupled to GM and RCs}

The nearly flat behavior of spiral galaxy RCs implies the
existence of dark matter with the energy density going like
$1/r^2$. This could be generated by a cylindrical halo around the
galaxy in non-relativistic Newton's theory of gravity. Since the
the rotation velocity in Eq. (2) can be written as $v^2_{rot}
\simeq -\vec{r}/2\cdot\nabla\tilde{g}_{tt} $ with $
\nabla\tilde{g}_{tt} =-2\hat{r}( (v_{rot}^{(0)})^2/r +
GM_{in}(r)/r^2 ) $, for instance, the singular isothermal sphere
\cite{Book} with energy density proportional to $1/r^2$ yields a
mass $M_{in} (r) \propto r$ and the third term in Eq. (1) gives us
a classical potential $\Psi \propto {\rm ln} (r/r_i)$. It
satisfies
\begin{equation}
\nabla^2 \Psi(r) = ( r \Psi')'/{r}= \Psi''+ \Psi'/{r} =0,
\end{equation}
with $ \Psi' \equiv \frac{d\Psi}{d r}$, ... in cylindrical
coordinates in flat space.

Such a cylindrical halo seems not to be realistic, but topological
defects such as GMs or cosmic strings \cite{string} have energy
densities proportional to $1/r^2$ and could be generated when
symmetry-breaking phase transitions took place in the early
Universe. The defects were thought of as the seeds for galaxy and
large-scale structure formation, and their remnants might remain
as galactic dark matter.
Though Harari and Loust\'{o} \cite{ha} proposed that the monopole
core mass is negative and that there exist no bound orbits in
Einstein's theory of gravity, Nucamendi and others \cite{nuc,bj}
suggested that the GM could account for some fraction of the
galactic dark matter.
It was also claimed by Nucamendi {\it et} {\it al.} \cite{nuc}
that there is an attractive region where bound orbits exist, by
the introduction of a nonminimal coupling of gravity to the GM,
$-\xi  R{\vec{\Phi}}^2$, which seems to play a role similar to the
interaction term $A^4 (\varphi){\vec{\Phi}}^2$ between the
monopole and BD field in Eq. (3) of Ref. \cite{TB}. We thus
suggested \cite{TB} that, in relativistic scalar-tensor theories
of gravity such as BD theory, GMs with the energy density $\propto
1/r^2$ can yield the logarithmic gravitational potential as the
second term in Eq. (1). This can be responsible for flat RCs,
while GMs induce only deficit angles in minimal coupling Einstein
gravity. It can be explicitly understood as follows.

In the spherically symmetric, static spacetime as $ds^2 \equiv
g_{\mu \nu} dX^\mu  dX^\nu
 =- b (r) c(r) dt^2 + dr^2 /b (r)
+r^2 d\Omega^2$, a $(t, t)$-component of the Einstein's equation
$G_{\mu}^{\nu} =\kappa T_{\mu}^{\nu}$ leads, upto $\cal O
(\kappa)$,
\begin{eqnarray}
b=1+\kappa b_1=1-\kappa \{ \eta^2(1+ {\delta^2}/{ r^2})+
{2M}/{r}\}+{\cal O}({\delta^4}/{ r^4} ),\nonumber
\end{eqnarray}
where $G_t^t =-(1-b)/r^2 + b' /r$ is used. The energy density of
the GM, $\rho (r) \equiv -T_t^t$, varies as $1/r^2$, for the
vacuum solution \cite{TB} $f_0=\eta \big( 1-\delta^2
/{r^2}\big)+{\cal O}(\delta^4/ r^4)$ of the monopole field
$\vec{\Phi} =f(r) \hat{r}$, when we consider the potential $ V_M
({\vec{\Phi}}^2 )=\frac{\lambda}{4}({\vec{\Phi}}^2 -\eta^2  )^2 $
with a constant $ \eta$ and the monopole core size
$\delta=1/{\sqrt{\lambda}\eta}$ \cite{vi}. Since a $(r,
r)$-component of the Einstein's equation gives us $c=1+{\cal O}
(\delta^4 /r^4)$ upto $\cal O (\kappa)$, there is no logarithmic
potential term generated by the GM in Einstein's theory of
gravity.

On the other hand, a generalized Einstein's equation in the
physical frame \cite{Bezerra} in BD gravity theory is given by
$\kappa T_t^t =G_t^t + b(\varphi')^2$, and thus the solution of BD
field is $\varphi=const +\kappa \varphi_1$ with $\varphi_1 =
 \alpha_0 (\eta^2  {\rm ln}(r/{r_i}) - M/{8 \pi
r})+{\cal O}(\delta^4/{ r^4}) $ for $\rho (r) \propto 1/r^2$
\cite{TB}. Since $\tilde{g}_{tt}=A^2 (\varphi) g_{tt} =
1/(G_*\tilde{\varphi})g_{tt}$, we have the gravitational potential
$\tilde{g}_{tt}$ proportional to ${\rm ln} (r/r_i)$. The massless
BD field satisfies also the field equation,
\begin{equation}
\varphi_1 '' + 2 \varphi_1 '/r =\alpha_0 \{ {f_0^2}/{r^2}+{1}/{2}(
f_0 ')^2  +2V_M (f_0) \},
\end{equation}
which can be approximated by a cylindrically symmetric equation
like Eq. (3), $\varphi_1 '' + \varphi_1 '/r \simeq 0$, since the
second term in the left hand side of Eq. (4) is twice as large as
the first term in the right hand side of the equation and
remaining terms are negligible in the large $r$ limit. The
effectively 2-dimensional structure formed due to topological
defects including GM (as candidates for dark matter) can be well
represented as a ${\rm ln}(r/r_i )$-like potential in BD theory of
gravity, while it is rather obscure in Einstein's theory of
gravity \cite{2D}. Other authors \cite{MOND} also found similar
gravitational potentials in various theories of gravity. Note that
the constant value $v_{rot.}^{(0)}$ originates from the
logarithmic term of the massless BD field contribution to
$\tilde{g}_{tt}$.

%%%%%%%%%%%%%%%%%%%%%%%%%%%%%%%%%%%%%%%%%%%%%%%%%%%%%%%%%%%%%%%%%%%%%%

Our model can not explain all detailed data of various spiral
galaxies, but it must be the basic framework on which we will
proceed to perform more realistic model building. Salucci and
Gentile's Comment \cite{Salucci} will be helpful for the purpose,
though we would like to notice a few things not to be
misinterpreted.

Firstly, they claimed that in our model the dark matter phenomenon
always emerges at outer radius $r$  of  a galaxy  as a constant
threshold value below which the circular velocity $v_{rot}(r)$
cannot decrease, regardless of the distance between $r$ and the
location  of the bulk  of  the stellar component. As we discussed
previously \cite{TB}, however, the formula (2) is valid only for
the finite range $r_i <r<r_h $ given by the range of global
monopole field \cite{nuc}, because the GM field (and BD field)
shall vanish at distance larger than the galactic halo radius
$r_h$ due to interactions with the nearest topological defect such
as antimonopole \cite{seed}, that is GM field lines can be
absorbed into the antimonopole core.
We moreover took a weak gravity approximation valid for $r<r_h <<
r_i e^{10^6} $ \cite{Ref}, and to go beyond the weak gravity
approximation we need a numerical analysis of the GM field coupled
to scalar-tensor gravity as Nucamendi {\it et al.} \cite{nuc}.

Secondly, they claimed that we predicted a constant value of
$v_{rot}^{(0)}= \sqrt{ 8\pi G \eta^2 \alpha_0^2/(1+\alpha_0^2)}
\sim 300\ km/s $, which is in disagreement with their result that
the extrapolated asymptotic amplitude $v(\infty)$ varies,
according to the galaxy luminosity, between $50\  km/s $ and $
250\ km/s $ \cite{URC}. From measurements of the RCs in spiral
galaxies we estimated the asymptotic value of $v_{rot}$ as between
$100\ km/s$ and $ 300\ km/s $ \cite{FRC,nuc}, for $ \eta \sim 3
\times 10^{16} {\rm GeV}$ and $10^{17} {\rm GeV} $, respectively,
which are the natural scales for grand unified theories(GUT)
\cite{vi,nuc}. Since we used an astronomical constraint for
$\alpha_0$ \cite{Nord}, $\alpha_0^2 \lesssim 0.001$, such that the
constraint was saturated, our value of $v_{rot}^{(0)}$ was not in
conflict with to Salucci and Gentile's \cite{Salucci}.

If the unsaturated constraint for $\alpha_0$ or a more stringent
bound on the value $\alpha_0$ given by experiments around a few AU
range \cite{Fisch} is used, then we will get a smaller value of
$v_{rot}^{(0)}$ (for instance, $v_{rot}^{(0)} \lesssim 100\ km/s$)
for such $\alpha_0$ and a fixed $\eta$. To fit various RC data of
spiral galaxies, we need more contributions from other (dark)
matter, $\Delta_{other}$, besides GM contributions,
and in this case Eq. (1) leads to
\begin{equation} \tilde{g}_{tt}(r)
\leadsto  -1- 2 \{(v_{rot}^{(0)})^2 +\Delta_{other} \}{\rm ln}
({r}/{r_i})+\cdots.
\end{equation}
Our model explains, as a universal property, the 'bare' flat
tendency of RCs in spiral galaxies which might be seeded by GMs
formed at the GUT scale $\eta$. To comprehend the detailed
'dressed' RCs of specific galaxies, we need other contributions
$\Delta_{other}$. With those contributions added to the GM
effects, ours can be phenomenologically more improved model of RCs
dependent on some properties of spiral galaxies. For example, the
potential of scalar fields considered by Matos and others
\cite{scalar} to account for the flat RCs could be adjusted
differently for different galaxies, which seems to us to play a
subsidiary role, as part of $\Delta_{other}$, though their models
themselves might be too specific as noted by Nucamendi {\it et
al.} \cite{nuc}.

The last thing is as follows. From available data of a sample of
about 1000 galaxies, Persic and others obtained the expression for
the universal RC \cite{URC,Rhee}, for a galaxy of luminosity
$L/L_*$ and normalized radius $x$,
\begin{equation}
v_{urc}^2(x)=G \big( M_h(1) F(x,L) + k { M_{in}(x)}/{x} \big),
\end{equation}
where $M_h(1)$ is the halo mass inside optical radii $R_{opt}$ and
$k$ of the order of unity. Then, they claimed that the dark
contribution $F(x,L)$ to the RC varies as $x^2/(x^2+a^2)$ with a
constant $a$ in each object, differently from our model and the
flat RC paradigm. Even though it was discussed in Ref.
\cite{Sofue} for Verheijen \cite{V} to show that one third of 30
spiral galaxies in the Ursa Major cluster have velocity curves
which do not conform to the universal curve shape and several
studies \cite{Tem} discussed the inadequacy of the universal RC
parametrization, we are going to make our model more applicable to
diverse RC models such as Persic {\it et al}'s \cite{URC} and to
study the cases with cored distributions of dark matter
\cite{Iliev}, $\rho (r) \propto ( {r_i}^2 + r^2)^{-m} r^{-n}$,
(with the core radius $r_i$ and constants $m$ and $n$) to be valid
for small $r$, since we can write the energy density of GM as
$\rho (r) = \eta^2(\delta^2+r^2)^{-1} $ upto ${\cal O} (\delta^4/{
r^4})$. In so doing, our formula (2) will be viable as a first
approximation ($a \rightarrow 0$) of various improved models
including the universal RC expressed as in Eq. (6).

%%%%%%%%%%%%%%%%%%%%%%%%%%%%%%%%%%%%%%%%%%%%%%%%%%%%%%%%%%%%%%%%%%%%%%%%
\section{Summary and discussion}

Though a GM induces only a deficit angle \cite{vi,ha} in
Einstein's theory of gravity, its energy density proportional to
$1/r^2$ generates the ${\rm ln}(r/r_i )$-term in a metric
component $\tilde{g}_{tt}$ in the physical frame \cite{Bezerra} in
BD theory. The logarithmic potential yields the constant value
$v_{rot}^{(0)}$ in the rotation velocity. The flat tendency of RCs
can be related with such a effectively 2-dimensional structure
formed because of GM, as discussed in the former part of the
previous section. If the flat part of RCs is really generated due
to GMs gravitationally coupled to the galaxies, then they would be
yielded in generalized theories of gravity (such as BD theory,
Dilaton gravity theories and nonminimal gravity theories) rather
than in Einstein's theory of gravity. We thus anticipate being
able to differentiate generalized theories of gravity from
Einstein's, by investigating thoroughly the motion of particles at
the galactic level through RC data.

The universal RC model \cite{URC} in spiral galaxies looks like an
improved one in comparison with the flat RC. However it was
discussed \cite{V, Tem} that the universal curve shape is not
confirmed yet and the universal RC parametrization may not be
adequate. Therefore the paradigm of flat RCs seems not to be
superseded by the universal RC as Salucci and Gentile
\cite{Salucci} argued, and we think that the notion of flat RCs
remains still as a foundation of various detailed studies on the
spiral galaxy RCs, which is discussed in the latter part of the
previous section.

\acknowledgments

We thank Profs. Myungshin Im, Soon-Tae Hong, and Jin Min Kim for
useful discussions. We also thank the referee for helpful
comments. This work was supported by the Soongsil University
Research Fund.

\end{document}